\UseRawInputEncoding 
\documentclass[
  draft
]{agujournal2025}
\usepackage{amsmath}
\usepackage{gensymb}
\begin{document}

\journalname{Geophysical Research Letters}

% Your title can be multiple lines long.
\title{Do AI Forecast Ensembles Sample the Correct Conditional Distribution?}

\authors{%
                  Lucas J. Howard\affil{1},
	          % Repeat the above for each author, with commas between authors, e.g.:
                  Elizabeth A. Barnes\affil{1}
	      }

% Repeat for all authors:	      
\affiliation{1}{Boston University, Faculty of Computing and Data Sciences} % First argument is their footnote number, as above.

% Corresponding author. Do not prepend with ``SI Corresponding author: '' in published version.
\authoraddr{%
                      % e.g., their full name, their department, their academic institution, their building, their City, State, Zip, Country. (theiremail\@ their institution)
                      Lucas Howard, Ljhoward@bu.edu.
                   }

\authorrunninghead{Ignored.} % Ignored.

% Key points.  
%   Takes three arguments.  Remember to specifiy them all, using {} for no item.
%
\keypoints%
    % First key point, etc.
{A trained probabilistic AI S2S coastal SSH forecast has positive marginal skill but negative joint skill.}
{The skill gap persists and is insensitive to training dataset volume in idealized experiments.}
{A deterministic emulator reproduces the failure mode while a dynamical model does not.}

% Call after \authors.

% Title running heads will be ignored, too:
\titlerunninghead{Ignored.}

% Key points was here

\maketitle

% Two options for abstracts: one with an abstract only and one with a plain language summary included.
%

\bigskip  % Remove; for demo purposes only.
% OR:
\begin{abstract}
Ensemble forecasting aims to sample the conditional distribution of outcomes; whether AI forecast ensembles do this correctly in a joint sense remains largely untested. We train a diffusion model for probabilistic subseasonal coastal sea level forecasts at eight US East Coast tide gauge stations, with sea level derived from reanalysis, and find that marginal and joint forecast quality decouple: positive skill at every station and lead time marginally, while joint spatial structure is worse than climatological draws. A shuffle-based permutation decomposition reveals this failure is invisible to the energy score but detected by the variogram score. Lorenz-96 experiments across 0.7-170 equivalent years show the gap persists regardless of training volume and is reproduced by a linear baseline, indicating structural inadequacy of the learned distribution. A dynamical ensemble does not replicate the failure while a deterministic emulator does, suggesting it is specific to learned emulators rather than ensemble forecasting generally.
\end{abstract}

\begin{plainlanguagesummary}
Forecasts of Earth's oceans and atmosphere are inherently uncertain. These systems are chaotic, our estimates of their current state are imprecise, and our models' representations of the underlying physics are imperfect. Reliably quantifying the uncertainty of a prediction is therefore a key component of modern forecasting. Ensemble forecasts, in which multiple plausible predictions are independently generated, are a common tool for this purpose. The rapid advancement of AI forecast models, which require far fewer computational resources than traditional models, allows for much larger ensembles and potentially better uncertainty estimates.

But a larger ensemble is only useful if it samples the right distribution, i.e. if its spread honestly reflects true uncertainty. We train an AI model to forecast weekly sea levels at eight tide gauge stations along the US East Coast and find that while the model produces skillful forecasts at each individual station, its predictions of how stations are correlated with one another are worse than random historical draws. Experiments in a simplified representation of the atmosphere show that adding more training data does not fix this gap, and that it appears in AI-based models but not traditional physics-based models suggesting a structural limitation of data-driven forecasting.
\end{plainlanguagesummary}

\section{Introduction}
The fundamental problem that ensemble forecasting attempts to (approximately) solve is to generate samples from a conditional distribution that represents the true uncertainty in outcomes \cite{leutbecher_ensemble_2008}. Examples of conditions could be the initial state, observations, or climate mode -- but in all cases the final ensemble ideally contains information about the ways and extent to which forecasts are constrained by available information \cite{gneiting_probabilistic_2014}. With ensemble size generally limited by available computational resources, the proliferation of efficient AI forecast tools offers the potential of much larger ensembles and higher fidelity estimates of uncertainty \cite{price_gencast_2023,weyn_sub-seasonal_2021}. In this work, we explore a foundational question for generative AI forecast ensembles, namely: what distribution are they truly representing, and how does it relate to actual uncertainty?

Earth system forecast models provide specific and actionable information to stakeholders on time scales ranging from weather to climate \cite{bauer_quiet_2015}. They undergird early warning systems that save lives and protect property, they inform guidance for key sectors of the global economy including transportation and agriculture, and they project future climate scenarios to support policy development and implementation. These models have traditionally been numerical solutions of the relevant governing equations solved on a discrete grid, which on the spatial and temporal scales involved are highly computationally expensive \cite{palmer_ecmwf_2019}. Rigorous uncertainty estimation via ensembles has therefore been challenging, with ensemble sizes limited by the cost of a single deterministic simulation. Rapid advances in the use of Artificial Intelligence (AI) methods in earth system science have changed this. Full AI weather emulators and hybrid systems in which only subgrid processes are AI-based have both demonstrated impressive forecast accuracy, competitive with or even surpassing traditional models on commonly used benchmarks \cite{bi_panguweather_2022,lam_graphcast_2022,price_gencast_2023}. Operational forecast agencies now run AI forecast models in parallel with their traditional dynamical models \cite{lang_aifs_2024}. Even as AI emulators have demonstrated impressive skill \cite{bi_panguweather_2022,lam_graphcast_2022,price_gencast_2023,lang_aifs_2024}, the underlying question of whether ensemble members are sampled from the correct joint distribution remains open.

In this work, we show that a generative model trained using reanalysis to predict weekly coastal sea surface height anomalies has positive marginal skill while simultaneously failing to capture the joint structure of the true uncertainty. Due to the limited size of the reanalysis dataset used for training, we use an idealized system to isolate the impact of training data volume on the observed gap between CRPS skill and variogram score skill. The results of this experiment show that this gap does not close with increasing training data volume out to an equivalent of 2X the ERA5 record, suggesting the failure is due to structural inadequacy of the distribution represented by the model’s output rather than insufficient training data. A dynamical ensemble does not replicate the failure mode, while a deterministic emulator using identical initial conditions does, suggesting that it is a vulnerability of data-driven approaches not shared by dynamical forecasts.

\section{Methods}
\subsection{Marginal and Joint Forecast Evaluation}
\label{sec:methods_eval}
To assess whether forecast ensembles are representing the correct distribution in both a marginal and joint sense, we use several evaluation metrics including the continuous ranked probability score (CRPS) \cite{gneiting_strictly_2007}, energy score (ES) \cite{gneiting_probabilistic_2014}, and variogram score (VS) \cite{scheuerer_variogram-based_2015}. CRPS is a proper scoring rule that assesses the marginal distribution of a forecast ensemble at a single location and lead time; a forecast minimizing CRPS is not required to correctly represent the joint distribution across locations. ES is a strictly proper multivariate extension of CRPS that in principle penalizes joint as well as marginal errors, but has been shown to be relatively insensitive to joint contributions when marginal skill dominates \cite{scheuerer_variogram-based_2015}. VS is an alternative multivariate proper scoring rule (not derived from CRPS) designed to be more sensitive to the dependence structure of a forecast. We use VS with power parameter p=0.5 and equal weights, as recommended by \citeA{scheuerer_variogram-based_2015} as a convenient default. All scores are expressed as skill scores of the form $1-S_forecast/S_ref$, where $S_forecast$ denotes the mean score over the evaluation period and $S_ref$ is the score of a climatological reference forecast (described in Section 2.2).

To isolate the contribution of the joint spatial structure to each score, we apply a shuffle-based permutation decomposition. For a given forecast ensemble, the joint structure is destroyed by independently permuting the ensemble values at each station, which preserves each station’s marginal distribution while eliminating cross-station correlations. We apply this shuffle separately to the forecast ensemble and to the climatological reference ensemble. The joint contribution is then quantified as

\begin{equation}
    \Delta = S(\text{original}) - S(\text{shuffled}),
\end{equation}

where a larger $\Delta$ indicates greater sensitivity of the score to the joint forecast structure. Comparing $\Delta_{forecast}$ to $\Delta_{ref}$ isolates the additional joint contribution of the DDPM above what climatology provides under the same score.

\subsection{Diffusion Model SSH Forecast}
\label{sec:ssh}
We train a denoising diffusion probabilistic model (DDPM) \cite{ho_denoising_2020} to generate probabilistic SSH anomaly forecasts at eight NOAA National Water Level Observation Network tide gauge stations along the US East Coast: Eastport ME, Boston MA, Atlantic City NJ, The Battery NY, Sewells Point VA, Charleston SC, Mayport FL, and Key West FL. The model outputs the SSH anomaly simultaneously at all eight locations, so the spatial joint distribution is determined entirely by the learned generative model.

\subsubsection*{Data and Preprocessing}
Ocean state at initialization is represented using GLORYS SSH and sea surface temperature (SST) over the western North Atlantic (5-50\degree N, 98-50\degree W), resampled to weekly means \cite{lellouche_copernicus_2021}. Atmospheric variability is represented by the North Atlantic Oscillation (NAO) index—defined as the leading principal component of ERA5 mean sea level pressure over 20 -80 N, 90 W-40 E—and by the 10-m horizontal wind components from ERA5 over the same domain \cite{kenigson_decadal_2018,thompson_annular_2000}.  A Ni\~no3.4 index (mean SST anomaly in 5\degree S-5\degree N, 170-120\degree W) derived from GLORYS is included to utilize potential ENSO teleconnections \cite{arcodia_subseasonal_2024}. All fields are converted to anomalies by a) subtracting the weekly mean computed over the training period (1993-2018) and b) removing the seasonal cycle. Spatial variability in SSH, SST, and wind is compressed using empirical orthogonal functions (EOFs) computed on the training period; we retain the 15 leading SSH modes, 15 SST modes, and 10 wind modes as predictors. SST and SSH components each explain 75\% of the variance in the dataset while wind components explain 71\%.

\subsubsection*{Model}
The DDPM conditioning vector is 131-dimensional and includes: the 15 leading SSH principal components at four initialization lags (weeks 0,-1,-2,-4); the 15 leading SST principal components at three lags (weeks 0, -1,-2); the 10 leading wind principal components at initialization; the NAO index at three lags (weeks 0,-1,-2); the Ni\~no 3.4 index at four lags (weeks 0, -4,-8,-13); the week 0 GLORYS SSH anomaly at the nearest valid cell to each of the 8 gauge locations; and the forecast lead time which, for training, is one of 2, 4, 8, or 12 weeks normalized to [0,1]. The denoising network is a three-hidden-layer fully connected neural network with layer widths (512, 512, 256). The diffusion process uses a linear noise schedule over T=1000 steps and ensemble members are generated by sampling the learned conditional distribution using the reverse diffusion process. Full training results and hyperparameter tuning can be found in Figure S1 and full architecture details in Table S1.

\subsubsection*{Training and Evaluation}
The model is trained on weekly mean initialization times from 1993-2018, validated on 2019-2021, and evaluated on a locked test set spanning 2022-2025 (184 initialization times). Training targets are GLORYS12v1 weekly mean SSH anomalies interpolated to each gauge location. An ensemble of 500 members is drawn independently from the conditional distribution for each test initialization at lead times of 2, 4, 8, and 12 weeks. The reference forecast for all skill scores is a climatological draw: 500 samples drawn uniformly from the training-period SSH anomaly pool, stratified by calendar week. Skill scores and confidence intervals are computed as described in Section \ref{sec:methods_eval}.

Confidence intervals on all skill scores and $\Delta$ values are obtained by a block bootstrap with block length 3 weeks (1000 resamples), which accounts for temporal autocorrelation between successive weekly initialization times.

\subsection{Lorenz-96 Experiments}
\label{sec:l96}
The Lorenz-96 (L96) system is a single-variable N-dimensional idealized system defined by N coupled ordinary differential equations with cyclic boundary conditions \cite{lorenz_designing_2005}. It is widely used as a simplified test system in earth system research \cite{arnold_stochastic_2013,brajard_combining_2020,gagne_machine_2020,howard_machine_2024}. We use it here to isolate the impact of training data volume and forecast model type on the CRPS/VS skill gap observed in the SSH experiments.
We use N=40 and forcing F=8 which are standard choices for which the system is known to be chaotic \cite{lorenz_designing_2005}. Based on the magnitude of the leading Lyapunov exponent, 1 model time unit (TU) is equivalent to approximately 5 Earth days.

The equations are integrated forward using a fourth-order Runge-Kutta scheme with time step dt=0.05 (20 steps per TU). After discarding 100 TU (2,000 steps) as spinup, we save a trajectory of 650,000 steps (32,500 TU). The first 250,000 steps form the training pool, followed by a 100-step gap, 100,000 validation steps, another 100-step gap, and 299,800 test steps. The gaps are included to ensure that each set is fully independent with no data leakage between training, testing, and validation. 500 test cases are drawn randomly without replacement for evaluation from the test set, ensuring that sampled states are well-separated in time and approximately independent.

A DDPM is trained to predict the full 40-dimensional system state at lead times of 20, 40, and 60 steps (1, 2, and 3 TU), conditioned on the initial state. Because the L96 domain is cyclic, the denoising network is a convolutional neural network with circular padding (4 residual blocks, 32 hidden channels). Training is repeated for eight training set sizes drawn from the training pool: 1,000; 5,000; 10,000; 20,000; 50,000; 100,000; 200,000; and 249,000 steps (approximately 0.7 to 170 equivalent years; the ERA5 record since 1940 corresponds to approximately 86 equivalent years). For each trained model, an ensemble of 20 members is generated for each of the 500 test initializations. Full architecture details can be found in Table S1.

As a linear baseline, we fit an ordinary least-squares regression with full residual covariance (multivariate normal, MVN) to each training set using the same conditioning (initial state) and evaluate it with CRPS and VS. Confidence intervals are obtained by paired bootstrap with 1,000 resamples of the 500 test initializations, applied identically to the DDPM and MVN.

We additionally compare the DDPM against two reference ensemble forecast types at lead=20 steps (1 TU). The first is a dynamical ensemble generated by integrating the L96 governing equations forward from perturbed initial conditions, where perturbations are drawn from an isotropic Gaussian with standard deviation $\sigma_{ic}$ reported in units normalized to the climatological standard deviation. The second is a deterministic emulator trained on the same L96 trajectories and with the same convolutional architecture as the DDPM. Ensemble members are generated by applying this fixed forward map to independently perturbed initial conditions. Both ensembles use 500 members and the same 500 test initializations as the DDPM. We evaluate each at two perturbation levels: $\sigma_{ic} = 0.01$ to minimize initial-condition uncertainty and isolate the effect of model structure, and a spread-matched $\sigma_{ic}$ calibrated by grid search to match the DDPM ensemble average spread at lead=20, making the absolute skill comparison fair.

\section{Results}
\subsection{Diffusion Model SSH Forecast}
\begin{figure}
    \centering
    \includegraphics[width=1\linewidth]{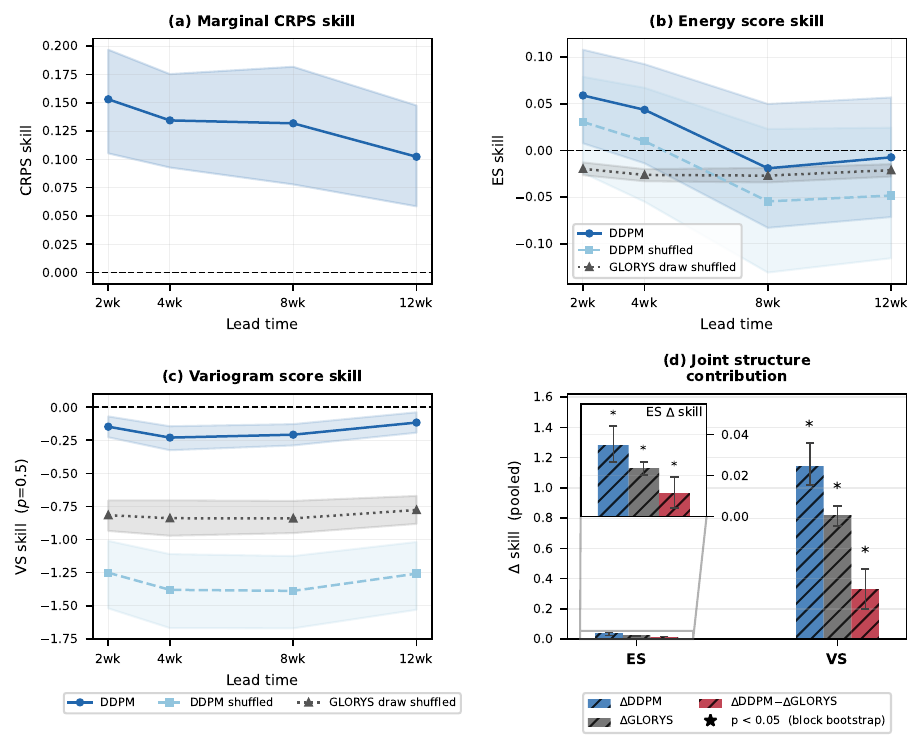}
    \caption{DDPM SSH forecast skill evaluated using three probabilistic metrics with 95\% confidence intervals indicated with shading and error bars. (a) CRPS skill score; positive values indicate improvement over a climatological forecast. (b) Energy Score (ES) skill score of the DDPM ensemble (filled circles), the spatially shuffled DDPM ensemble (open squares), and a shuffled GLORYS climatological draw (triangles) (c) As in (b) but for Variogram Score (VS). (d) Lead-pooled joint structure contribution ($\Delta$ skill) for ES and VS; inset shows ES on an expanded scale. The DDPM achieves positive CRPS and ES skill but negative VS skill at all leads, demonstrating that standard metrics can mask a substantial joint forecast failure.}
    \label{fig:figure1}
\end{figure}

The DDPM achieves statistically significant positive CRPS skill at all four lead times (Figure \ref{fig:figure1}a), with station-mean skill dropping monotonically with increasing lead time. Energy Score skill is positive for short leads and slightly negative for longer leads (Figure \ref{fig:figure1}b). The negative skill at 8 and 12 weeks is not statistically significant. In contrast, Variogram Score skill is negative at all lead times (Figure \ref{fig:figure1}c), indicating that the joint spatial structure of the DDPM ensemble is worse than a climatological draw despite the positive marginal skill.

The shuffle decomposition in Figure \ref{fig:figure1}b)-(d) isolates the contribution of joint ensemble structure to each score. Figure \ref{fig:figure1}b shows the ES decomposition. The joint contribution of the DDPM is statistically significant (asterisk) but small and only modestly larger than the joint contribution of the climatological reference, i.e. the DDPM’s joint structure provides little additional information beyond what a random climatological draw provides. Figure \ref{fig:figure1}c shows the VS decomposition. Using VS, the joint contribution to DDPM skill is substantially and significantly larger than the skill of a climatological forecast at every lead despite the fact that overall skill is negative. Figure \ref{fig:figure1}d shows the lead-pooled joint contributions. Under ES, the DDPM's joint contribution barely exceeds that of a climatological draw -- the red bar is near zero. Under VS, the DDPM's joint contribution substantially exceeds climatology, yet overall VS skill remains negative (Figure \ref{fig:figure1}c), indicating that the DDPM produces joint structure that is detectably non-climatological but still worse than the observed joint distribution. ES and VS detect fundamentally different aspects of the joint forecast distribution: ES is dominated by marginal contributions and is largely insensitive to the multivariate structure of the forecast, while VS clearly detects a structural failure of the distribution represented by the ensemble. 

\begin{figure}
    \centering
    \includegraphics[width=1\linewidth]{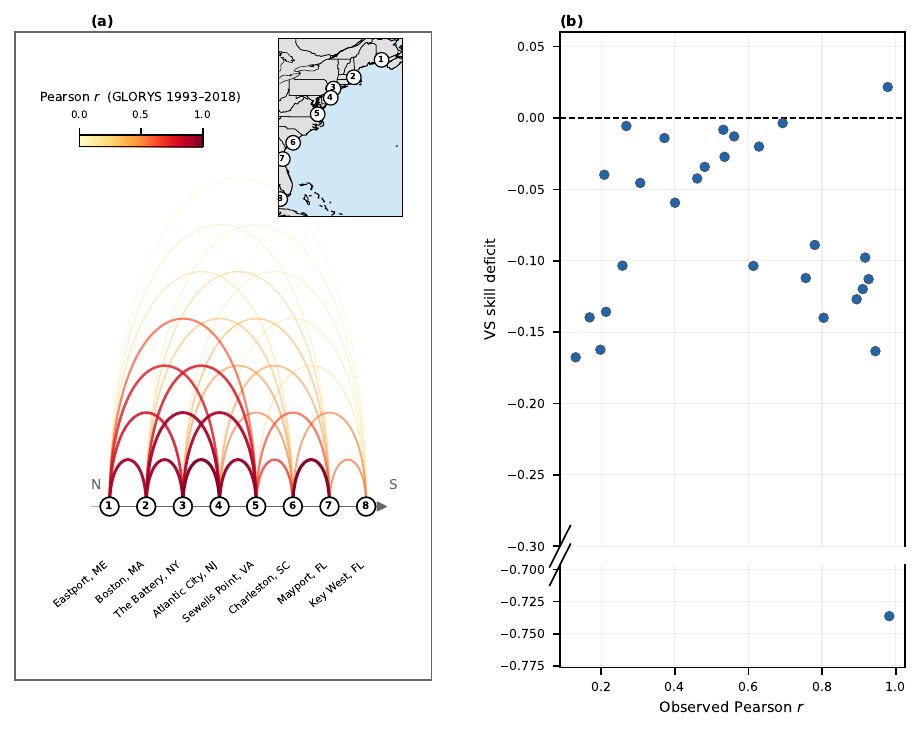}
    \caption{Pairwise SSH correlation structure and per-pair VS skill deficit across the eight tide gauge stations. (a) Arc diagram of pairwise Pearson r between stations ordered north (left) to south (right); arc color and thickness indicate correlation strength (GLORYS 1993-2018 training period). Inset shows station locations along the US East Coast. (b) The multivariate skill deficit vs. the observed correlation between stations. Skill deficit is highly heterogeneous across station correlation strength.}
    \label{fig:figure2}
\end{figure}

To understand this result within the context of the underlying correlation structure of the system, we next examine the pairwise correlations from the training set (Figure \ref{fig:figure2}a) as well as the magnitude of the negative VS skill for that pair averaged over the test set (Figure \ref{fig:figure2}b). There are strong observed correlations concentrated among geographically adjacent northern stations (Eastport through Sewells Point) and among the two southernmost stations (Mayport and Key West). Most station pairs have negative VS skill, indicating the DDPM ensemble is worse than climatology for nearly all pairs. The skill deficit is strongly heterogeneous: the Atlantic City-Battery pair is an extreme outlier, with the highest observed correlation (r=0.98) and the largest VS deficit, consistent with the DDPM over-correlating this already highly correlated pair. Southern station pairs (Mayport, Key West) show smaller deficits and, in several cases, near-zero failure, hinting that large scale drivers of joint (i.e. spatially correlated) variability may be better captured at these locations. This heterogeneity suggests that the failure mode is not caused by the DDPM simply uniformly amplifying or suppressing correlation structure. Straightforward correction using postprocessing is therefore non-trivial.

\subsection{Lorenz 96 Results}
\subsubsection{Sensitivity to Training Volume}
\begin{figure}
    \centering
    \includegraphics[width=1\linewidth]{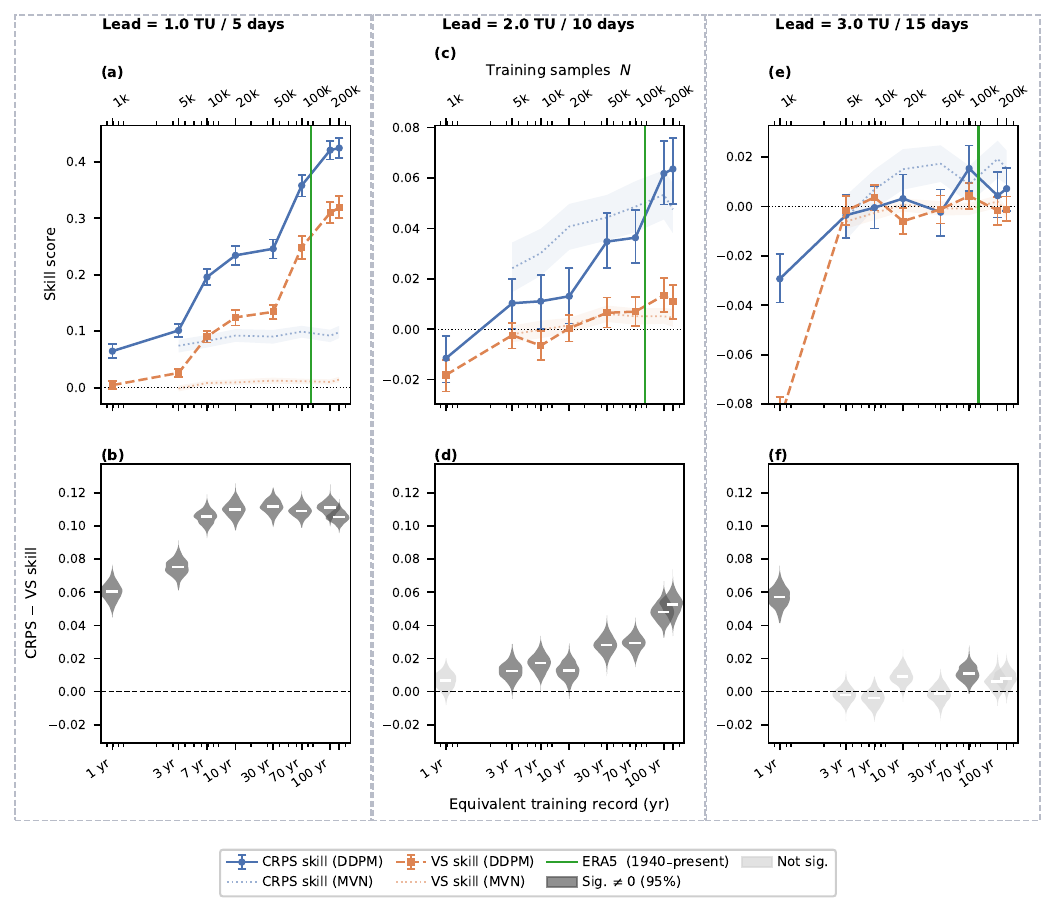}
    \caption{CRPS and Variogram Score (VS) skill as a function of training data volume for a DDPM trained on the Lorenz-96 system at three forecast lead times. (a, c, e) CRPS skill (blue) and VS skill (orange) scores with 95\% confidence intervals as a function of training period length (x-axis, in equivalent years) for lead times of 5, 10, and 15 days, respectively. The ERA5 equivalent record length is indicated by a solid green vertical line; the skill of a multivariate normal (MVN) baseline is shown for comparison (dotted line). (b, d, f) Violin plots show distributions of the CRPS-VS skill difference at each training set size for the same lead times; dark fill indicates the gap is significantly different from zero (95\% bootstrap CI). A statistically significant positive gap ($\text{CRPS}>\text{VS}$) is present at leads of 5 and 10 days across all training volumes and shows no systematic decline as training data increases to 2$\times$ the ERA5 equivalent record, indicating that the marginal-joint skill discrepancy is not resolved by additional training data under the conditions tested.}
    \label{fig:figure3}
\end{figure}

We next explore the Lorenz-96 experiments to determine dependence on training data volume. At a lead of 5 days (1 TU, approximately 1.7 Lyapunov times), both CRPS and VS skill increase with training volume, but CRPS skill is consistently higher than VS skill and a statistically significant gap between CRPS and VS skill is present at all training volumes tested (Figure \ref{fig:figure3}a,b). At a lead of 10 days (2 TU, approximately 3.3 Lyapunov times), CRPS skill continues to grow with training volume while VS skill remains near zero across all training set sizes, demonstrating a near-complete decoupling of marginal and joint forecast improvement (Figure \ref{fig:figure3}c,d). At 15 days (3 TU, approximately 5.0 Lyapunov times), both scores are near zero at all training volumes, consistent with this lead exceeding the effective predictability horizon of the system (Figure \ref{fig:figure3}e,f).

The MVN baseline—an OLS regression with full residual covariance fit to the same training data—exhibits the same qualitative pattern at every lead and training volume: CRPS skill tracks closely with the DDPM while VS skill remains near zero or below. The absence of learnable joint signal is not specific to the nonlinear DDPM architecture; the much simpler linear regression recovers the same marginal skill compared to climatology without recovering joint calibration with increasing data volume. The CRPS-VS decoupling observed in the SSH forecast therefore should not be dismissed as a consequence of insufficient training data.

\subsubsection{Dynamical Model and Deterministic AI Emulator}
To identify the structural source of the CRPS-VS gap, we compare the DDPM against dynamical and deterministic emulator ensembles at lead=20 steps (1 TU, Figure 4). Additional results for both forecasts are shown in Figure S2 and S3. In the quasi-linear dynamical regime ($\sigma_{ic}$ = 0.01), the VS/CRPS ratio is approximately 1.0 (ratio = 1.02), confirming that chaotic divergence alone does not produce a joint calibration deficit. When $\sigma_{ic}$ is tuned to match the DDPM ensemble spread ($\sigma_{ic}$ = 0.158), the dynamical ratio falls modestly to 0.91, indicating that nonlinear trajectory divergence introduces a small joint deficit but cannot account for the full gap observed in the DDPM (ratio = 0.75). The deterministic emulator performs worst at both noise levels (VS/CRPS ≈ 0.42-0.58), consistent with a deterministic forward map collapsing ensemble diversity in a way that disproportionately harms joint relative to marginal calibration. Both AI-based methods show a substantially lower VS/CRPS ratio than the dynamical ensemble at matched spread, suggesting the joint calibration gap is a property of data-driven emulation rather than a consequence of ensemble spread or nonlinear divergence alone.

\begin{figure}
    \centering
    \includegraphics[width=1\linewidth]{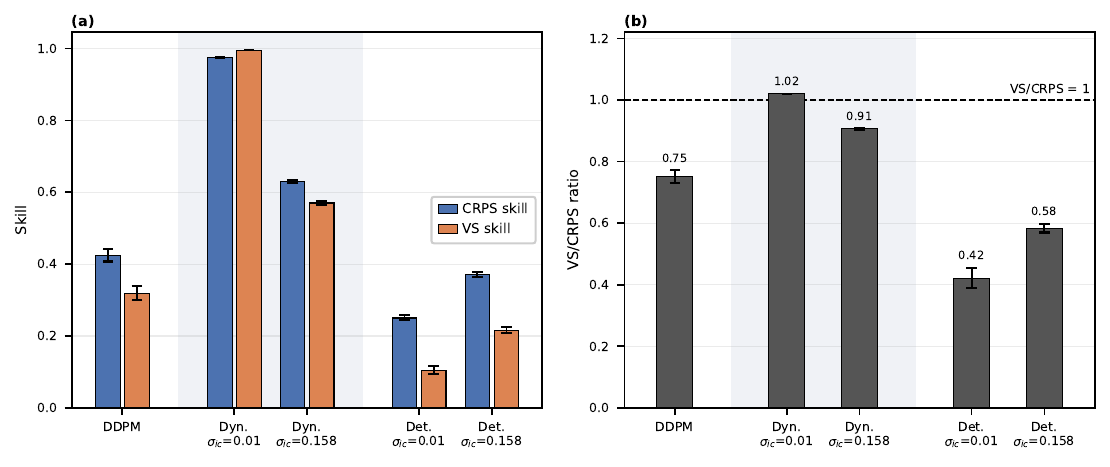}
    \caption{CRPS (blue) and VS (orange) skill scores for each of the ensemble modeling approaches (a). The DDPM is natively probabilistic, while the dynamical and deterministic emulators use two sets of perturbed initial conditions to generate ensemble forecasts. The first uses independently and identically distributed Gaussian noise with a standard deviation equal to 0.01 times the climatological standard deviation. The second uses a standard deviation tuned so that the dynamical and DDPM models have approximately equal average ensemble spread at the chosen forecast lead time of 1 TU (5 days). 95\% confidence intervals are included for all bars using block bootstrapping. (b) The ratio of VS to CRPS skill for all five experiments. A horizontal dashed line at y=1 corresponds to no gap in marginal and joint skill. The VS/CRPS gap is present in both AI-based methods at both noise levels but absent or substantially reduced in the dynamical forecast, indicating the failure is a property of data-driven emulation rather than ensemble forecasting in chaotic systems generally.}
    \label{fig:figure4}
\end{figure}

\section{Discussion and Conclusions}
We have shown that a diffusion model trained for subseasonal coastal sea level prediction achieves positive CRPS skill while producing joint spatial forecast structure that is systematically worse than climatology, and that energy score alone does not reveal this failure. A shuffle-based permutation decomposition exposes the discrepancy: variogram score skill is negative at all leads while energy score skill is positive, and the per-pair variogram score attribution shows highly spatially heterogeneous decoupling of marginal and joint skill. The joint contribution to ES skill is small and only modestly above what a climatological draw provides, while VS detects a substantially larger joint contribution, but one that is still miscalibrated. Lorenz-96 experiments across training volumes equivalent to 0.7-170 years of ERA5 demonstrate that the CRPS–variogram score gap does not close with additional data and is shared by linear and nonlinear forecast models alike; the MVN (which by design captures only the linear signal) exhibits the same insensitivity to training data quantity, establishing that the decoupling is unlikely to be a data-volume artifact.

Last, a direct comparison against ensembles produced with a dynamical model and a deterministic emulator using perturbed initial conditions shows the skill gap exists in the emulator but not in the dynamical model. Taken together, these experiments demonstrate that 1) probabilistic generative AI forecast tools may sample from a distribution that does not properly represent the joint conditional structure of the true uncertainty, 2) additional training data cannot be assumed to remedy this failure mode, 3) CRPS and ES used as evaluation or training metrics \cite{perkins_hiro-ace_2026} will not reliably identify or mitigate this failure mode, and 4) the failure mode is reproduced for a deterministic AI emulator but not a dynamical model.

Supporting the relevance of our findings, recent work on integrating AI forecast models into data assimilation systems has identified covariance errors as a fundamental limitation. \citeA{tian_evaluating_2026} showed that the tangent linear trajectories of deterministic AI models have exaggerated sensitivities and predicted that this would manifest as distorted background error covariances. \citeA{slivinski_assimilating_2025} confirmed this prediction empirically, finding that cycling EnKF with AI models produced covariance amplification sufficient to cause forecast divergence and instabilities. Our results are consistent with this overall picture of AI emulators producing ensembles that are miscalibrated in a multivariate sense even as they compare favorably on standard skill metrics.

AI forecast ensembles may appear skillful while simultaneously representing the multivariate distribution incorrectly with no guarantee this failure will improve with increased training data quantity. This strongly supports the continued use of dynamical and physics-based models to complement AI forecasts, and suggests that changes to both the practice and techniques of model evaluation and training are needed to fully realize the potential of AI forecast tools.

\section*{AI Disclosure}
Claude Code (Anthropic) was used to assist in generating the code for the experiments described in this project, as well as scripts for creating publication figures. AI tools, including Claude (Anthropic) and Chat GPT (OpenAI) were also used to assist with manuscript editing. The authors reviewed and verified the content of both the associated code repository and manuscript text and take responsibility for their accuracy and correctness.

\section*{Acknowledgments}
This work was supported, in part, by NOAA grants NA24OARX431C0022 and NA22OAR4310621. The computational work reported in this paper was performed using the Shared Computing Cluster which is administered by Boston University’s Research Computing Services.

% Your Data Availability Statement is an unnumbered section right before the bibliography.
\section*{Open Research Statement}
Raw GLORYS SSH and SST and ERA5 wind and SLP data are available from their originating agencies. All code for analysis, Lorenz 96 experiments, and plotting along with processed GLORYS and ERA5 data used in this paper are publicly available \cite{howard2026diffusion}.

% Bibliography
%\cite{*} % If you have uncited bibliography items.
%
\bibliography{references} % Replace ``wiley'' with your bibliography's file name.  Do not specify its extension, e.g., .bib.

@article{gneiting_strictly_2007,
    title = {Strictly {Proper} {Scoring} {Rules}, {Prediction}, and {Estimation}},
    volume = {102},
    issn = {0162-1459},
    url = {https://doi.org/10.1198/016214506000001437},
    doi = {10.1198/016214506000001437},
    number = {477},
    urldate = {2022-12-26},
    journal = {Journal of the American Statistical Association},
    publisher = {Taylor \& Francis},
    author = {Gneiting, Tilmann and Raftery, Adrian E},
    month = mar,
    year = {2007},
    note = {\_eprint: https://doi.org/10.1198/016214506000001437},
    pages = {359--378},
}

@article{howard_machine_2024,
    title = {A {Machine} {Learning} {Augmented} {Data} {Assimilation} {Method} for {High}-{Resolution} {Observations}},
    volume = {16},
    copyright = {© 2024 The Authors. Journal of Advances in Modeling Earth Systems published by Wiley Periodicals LLC on behalf of American Geophysical Union.},
    issn = {1942-2466},
    url = {https://onlinelibrary.wiley.com/doi/abs/10.1029/2023MS003774},
    doi = {10.1029/2023MS003774},
    language = {en},
    number = {1},
    urldate = {2024-01-31},
    journal = {Journal of Advances in Modeling Earth Systems},
    author = {Howard, Lucas J. and Subramanian, Aneesh and Hoteit, Ibrahim},
    year = {2024},
    note = {\_eprint: https://agupubs.onlinelibrary.wiley.com/doi/pdf/10.1029/2023MS003774},
    pages = {e2023MS003774},
}

@article{scheuerer_variogram-based_2015,
    chapter = {Monthly Weather Review},
    title = {Variogram-{Based} {Proper} {Scoring} {Rules} for {Probabilistic} {Forecasts} of {Multivariate} {Quantities}},
    volume = {143},
    issn = {1520-0493, 0027-0644},
    url = {https://journals.ametsoc.org/view/journals/mwre/143/4/mwr-d-14-00269.1.xml},
    doi = {10.1175/MWR-D-14-00269.1},
    language = {EN},
    number = {4},
    urldate = {2026-05-04},
    journal = {Monthly Weather Review},
    publisher = {American Meteorological Society},
    author = {Scheuerer, Michael and Hamill, Thomas M.},
    month = apr,
    year = {2015},
    pages = {1321--1334},
}

@article{brajard_combining_2020,
    title = {Combining data assimilation and machine learning to emulate a dynamical model from sparse and noisy observations: {A} case study with the {Lorenz} 96 model},
    volume = {44},
    issn = {1877-7503},
    shorttitle = {Combining data assimilation and machine learning to emulate a dynamical model from sparse and noisy observations},
    url = {https://www.sciencedirect.com/science/article/pii/S1877750320304725},
    doi = {10.1016/j.jocs.2020.101171},
    language = {en},
    urldate = {2022-10-11},
    journal = {Journal of Computational Science},
    author = {Brajard, Julien and Carrassi, Alberto and Bocquet, Marc and Bertino, Laurent},
    month = jul,
    year = {2020},
    pages = {101171},
}

@article{arnold_stochastic_2013,
    title = {Stochastic parametrizations and model uncertainty in the {Lorenz} ’96 system},
    volume = {371},
    url = {https://royalsocietypublishing.org/doi/10.1098/rsta.2011.0479},
    doi = {10.1098/rsta.2011.0479},
    number = {1991},
    urldate = {2021-10-12},
    journal = {Philosophical Transactions of the Royal Society A: Mathematical, Physical and Engineering Sciences},
    publisher = {Royal Society},
    author = {Arnold, H. M. and Moroz, I. M. and Palmer, T. N.},
    month = may,
    year = {2013},
    pages = {20110479},
}

@article{gagne_machine_2020,
    title = {Machine {Learning} for {Stochastic} {Parameterization}: {Generative} {Adversarial} {Networks} in the {Lorenz} '96 {Model}},
    volume = {12},
    issn = {1942-2466},
    shorttitle = {Machine {Learning} for {Stochastic} {Parameterization}},
    url = {https://agupubs.onlinelibrary.wiley.com/doi/abs/10.1029/2019MS001896},
    doi = {10.1029/2019MS001896},
    language = {en},
    number = {3},
    urldate = {2021-09-03},
    journal = {Journal of Advances in Modeling Earth Systems},
    author = {Gagne, David John and Christensen, Hannah M. and Subramanian, Aneesh C. and Monahan, Adam H.},
    year = {2020},
    note = {\_eprint: https://agupubs.onlinelibrary.wiley.com/doi/pdf/10.1029/2019MS001896},
    pages = {e2019MS001896},
}

@article{slivinski_assimilating_2025,
    title = {Assimilating {Observed} {Surface} {Pressure} {Into} {ML} {Weather} {Prediction} {Models}},
    volume = {52},
    copyright = {© 2025. The Author(s). This article has been contributed to by U.S. Government employees and their work is in the public domain in the USA.},
    issn = {1944-8007},
    url = {https://onlinelibrary.wiley.com/doi/abs/10.1029/2024GL114396},
    doi = {10.1029/2024GL114396},
    language = {en},
    number = {6},
    urldate = {2025-07-13},
    journal = {Geophysical Research Letters},
    author = {Slivinski, L. C. and Whitaker, J. S. and Frolov, S. and Smith, T. A. and Agarwal, N.},
    year = {2025},
    note = {\_eprint: https://agupubs.onlinelibrary.wiley.com/doi/pdf/10.1029/2024GL114396},
    pages = {e2024GL114396},
}

@article{lellouche_copernicus_2021,
    title = {The {Copernicus} {Global} 1/12° {Oceanic} and {Sea} {Ice} {GLORYS12} {Reanalysis}},
    volume = {9},
    issn = {2296-6463},
    url = {https://www.frontiersin.org/journals/earth-science/articles/10.3389/feart.2021.698876/full},
    doi = {10.3389/feart.2021.698876},
    language = {English},
    urldate = {2024-08-28},
    journal = {Frontiers in Earth Science},
    publisher = {Frontiers},
    author = {Lellouche, Jean-Michel and Greiner, Eric and Bourdallé-Badie, Romain and Garric, Gilles and Melet, Angélique and Drévillon, Marie and Bricaud, Clément and Hamon, Mathieu and Le Galloudec, Olivier and Regnier, Charly and Candela, Tony and Testut, Charles-Emmanuel and Gasparin, Florent and Ruggiero, Giovanni and Benkiran, Mounir and Drillet, Yann and Le Traon, Pierre-Yves},
    month = jul,
    year = {2021},
}

@misc{ho_denoising_2020,
    title = {Denoising {Diffusion} {Probabilistic} {Models}},
    url = {http://arxiv.org/abs/2006.11239},
    doi = {10.48550/arXiv.2006.11239},
    urldate = {2026-06-10},
    publisher = {arXiv},
    author = {Ho, Jonathan and Jain, Ajay and Abbeel, Pieter},
    month = dec,
    year = {2020},
    note = {arXiv:2006.11239 [cs.LG]},
}

@article{lorenz_designing_2005,
    chapter = {Journal of the Atmospheric Sciences},
    title = {Designing {Chaotic} {Models}},
    volume = {62},
    issn = {0022-4928, 1520-0469},
    url = {https://journals.ametsoc.org/view/journals/atsc/62/5/jas3430.1.xml},
    doi = {10.1175/JAS3430.1},
    language = {EN},
    number = {5},
    urldate = {2022-06-07},
    journal = {Journal of the Atmospheric Sciences},
    publisher = {American Meteorological Society},
    author = {Lorenz, Edward N.},
    month = may,
    year = {2005},
    pages = {1574--1587},
}

@article{tian_evaluating_2026,
    title = {Evaluating {Machine} {Learning} {Weather} {Models} for {Data} {Assimilation}: {Fundamental} {Limitations} in {Tangent} {Linear} and {Adjoint} {Properties}},
    volume = {53},
    copyright = {© 2026. The Author(s).},
    issn = {1944-8007},
    shorttitle = {Evaluating {Machine} {Learning} {Weather} {Models} for {Data} {Assimilation}},
    url = {https://onlinelibrary.wiley.com/doi/abs/10.1029/2025GL119402},
    doi = {10.1029/2025GL119402},
    language = {en},
    number = {2},
    urldate = {2026-06-10},
    journal = {Geophysical Research Letters},
    author = {Tian, Xiaoxu and Holdaway, Daniel and Kleist, Daryl},
    year = {2026},
    note = {\_eprint: https://agupubs.onlinelibrary.wiley.com/doi/pdf/10.1029/2025GL119402},
    pages = {e2025GL119402},
}

@article{bi_panguweather_2022,
    title = {Pangu‑{Weather}: {A} {3D} high‑resolution model for fast and accurate global weather forecast},
    journal = {arXiv preprint arXiv:2211.02556},
    author = {Bi, Kaifeng and Xie, Lingxi and Zhang, Hengheng and Chen, Xin and Gu, Xiaotao and Tian, Qi},
    year = {2022},
}

@article{lam_graphcast_2022,
    title = {{GraphCast}: {Learning} skillful medium-range global weather forecasting},
    journal = {arXiv preprint arXiv:2212.12794},
    author = {Lam, Rémi and Sanchez‑Gonzalez, Álvaro and Willson, Matthew and Wirnsberger, Peter and Fortunato, Meire and Alet, Ferran and Ravuri, Suman and Ewalds, Timo and Eaton‑Rosen, Zach and Hu, Weihua and Merose, Alexander and {others}},
    year = {2022},
}

@article{price_gencast_2023,
    title = {{GenCast}: {Diffusion}‑based ensemble forecasting for medium‑range weather},
    journal = {arXiv preprint arXiv:2312.15796},
    author = {Price, Ilan and Sanchez‑Gonzalez, Álvaro and Alet, Ferran and Andersson, Tom R. and El‑Kadi, Andrew and Masters, Dominic and Ewalds, Timo and Stott, Jacklynn and Mohamed, Shakir and Battaglia, Peter and {others}},
    year = {2023},
}

@misc{perkins_hiro-ace_2026,
    title = {{HiRO}-{ACE}: {Fast} and skillful {AI} emulation and downscaling trained on a 3 km global storm-resolving model},
    shorttitle = {{HiRO}-{ACE}},
    url = {http://arxiv.org/abs/2512.18224},
    doi = {10.48550/arXiv.2512.18224},
    urldate = {2026-08-03},
    publisher = {arXiv},
    author = {Perkins, W. Andre and Kwa, Anna and McGibbon, Jeremy and Arcomano, Troy and Clark, Spencer K. and Watt-Meyer, Oliver and Bretherton, Christopher S. and Harris, Lucas M.},
    month = feb,
    year = {2026},
    note = {arXiv:2512.18224 [physics.ao-ph]},
}

@article{bauer_quiet_2015,
    title = {The quiet revolution of numerical weather prediction},
    volume = {525},
    copyright = {2015 Nature Publishing Group, a division of Macmillan Publishers Limited. All Rights Reserved.},
    issn = {1476-4687},
    url = {https://www.nature.com/articles/nature14956},
    doi = {10.1038/nature14956},
    language = {en},
    number = {7567},
    urldate = {2022-12-15},
    journal = {Nature},
    publisher = {Nature Publishing Group},
    author = {Bauer, Peter and Thorpe, Alan and Brunet, Gilbert},
    month = sep,
    year = {2015},
    note = {Number: 7567},
    pages = {47--55},
}

@article{palmer_ecmwf_2019,
    title = {The {ECMWF} ensemble prediction system: {Looking} back (more than) 25 years and projecting forward 25 years},
    volume = {145},
    copyright = {© 2018 The Authors. Quarterly Journal of the Royal Meteorological Society published by John Wiley \& Sons Ltd on behalf of the Royal Meteorological Society.},
    issn = {1477-870X},
    shorttitle = {The {ECMWF} ensemble prediction system},
    url = {https://onlinelibrary.wiley.com/doi/abs/10.1002/qj.3383},
    doi = {10.1002/qj.3383},
    language = {en},
    number = {S1},
    urldate = {2026-08-03},
    journal = {Quarterly Journal of the Royal Meteorological Society},
    author = {Palmer, Tim},
    year = {2019},
    note = {\_eprint: https://rmets.onlinelibrary.wiley.com/doi/pdf/10.1002/qj.3383},
    pages = {12--24},
}

@article{arcodia_subseasonal_2024,
    chapter = {Weather and Forecasting},
    title = {Subseasonal {Variability} of {U}.{S}. {Coastal} {Sea} {Level} from {MJO} and {ENSO} {Teleconnection} {Interference}},
    volume = {39},
    issn = {1520-0434, 0882-8156},
    url = {https://journals.ametsoc.org/view/journals/wefo/39/2/WAF-D-23-0002.1.xml},
    doi = {10.1175/WAF-D-23-0002.1},
    language = {EN},
    number = {2},
    urldate = {2026-08-03},
    journal = {Weather and Forecasting},
    publisher = {American Meteorological Society},
    author = {Arcodia, Marybeth C. and Becker, Emily and Kirtman, Ben P.},
    month = feb,
    year = {2024},
    pages = {441--458},
}

@article{kenigson_decadal_2018,
    chapter = {Journal of Climate},
    title = {Decadal {Shift} of {NAO}-{Linked} {Interannual} {Sea} {Level} {Variability} along the {U}.{S}. {Northeast} {Coast}},
    volume = {31},
    issn = {0894-8755, 1520-0442},
    url = {https://journals.ametsoc.org/view/journals/clim/31/13/jcli-d-17-0403.1.xml},
    doi = {10.1175/JCLI-D-17-0403.1},
    language = {EN},
    number = {13},
    urldate = {2026-08-03},
    journal = {Journal of Climate},
    publisher = {American Meteorological Society},
    author = {Kenigson, Jessica S. and Han, Weiqing and Rajagopalan, Balaji and Yanto and Jasinski, Mike},
    month = jul,
    year = {2018},
    pages = {4981--4989},
}

@article{thompson_annular_2000,
    chapter = {Journal of Climate},
    title = {Annular {Modes} in the {Extratropical} {Circulation}. {Part} {I}: {Month}-to-{Month} {Variability}},
    volume = {13},
    issn = {0894-8755, 1520-0442},
    shorttitle = {Annular {Modes} in the {Extratropical} {Circulation}. {Part} {I}},
    url = {https://journals.ametsoc.org/view/journals/clim/13/5/1520-0442_2000_013_1000_amitec_2.0.co_2.xml},
    doi = {10.1175/1520-0442(2000)013<1000:AMITEC>2.0.CO;2},
    language = {EN},
    number = {5},
    urldate = {2026-08-03},
    journal = {Journal of Climate},
    publisher = {American Meteorological Society},
    author = {Thompson, David W. J. and Wallace, John M.},
    month = mar,
    year = {2000},
    pages = {1000--1016},
}

@article{leutbecher_ensemble_2008,
    series = {Predicting weather, climate and extreme events},
    title = {Ensemble forecasting},
    volume = {227},
    issn = {0021-9991},
    url = {https://www.sciencedirect.com/science/article/pii/S0021999107000812},
    doi = {10.1016/j.jcp.2007.02.014},
    language = {en},
    number = {7},
    urldate = {2023-02-23},
    journal = {Journal of Computational Physics},
    author = {Leutbecher, M. and Palmer, T. N.},
    month = mar,
    year = {2008},
    pages = {3515--3539},
}

@article{gneiting_probabilistic_2014,
    title = {Probabilistic {Forecasting}},
    volume = {1},
    url = {https://doi.org/10.1146/annurev-statistics-062713-085831},
    doi = {10.1146/annurev-statistics-062713-085831},
    number = {1},
    urldate = {2023-02-23},
    journal = {Annual Review of Statistics and Its Application},
    author = {Gneiting, Tilmann and Katzfuss, Matthias},
    year = {2014},
    note = {\_eprint: https://doi.org/10.1146/annurev-statistics-062713-085831},
    pages = {125--151},
}

@article{weyn_sub-seasonal_2021,
    title = {Sub-{Seasonal} {Forecasting} {With} a {Large} {Ensemble} of {Deep}-{Learning} {Weather} {Prediction} {Models}},
    volume = {13},
    copyright = {© 2021. The Authors. Journal of Advances in Modeling Earth Systems published by Wiley Periodicals LLC on behalf of American Geophysical Union.},
    issn = {1942-2466},
    url = {https://onlinelibrary.wiley.com/doi/abs/10.1029/2021MS002502},
    doi = {10.1029/2021MS002502},
    language = {en},
    number = {7},
    urldate = {2026-08-04},
    journal = {Journal of Advances in Modeling Earth Systems},
    author = {Weyn, Jonathan A. and Durran, Dale R. and Caruana, Rich and Cresswell-Clay, Nathaniel},
    year = {2021},
    pages = {e2021MS002502},
}

@article{lang_aifs_2024,
    title = {{AIFS} – {ECMWF}’s data‑driven forecasting system},
    journal = {arXiv preprint arXiv:2406.01465},
    author = {Lang, Simon and Alexe, Mihai and Chantry, Matthew and Dramsch, Jesper and Pinault, Florian and Raoult, Baudouin and Clare, Mariana C. A. and Lessig, Christian and Maier‑Gerber, Michael and Magnusson, Linus and {others}},
    year = {2024},
}

@misc{howard2026diffusion,
    title        = {Do {AI} Forecast Ensembles Sample the Correct Conditional Distribution? Code Repository},
    author       = {Howard, Lucas},
    year         = {2026},
    month        = aug,
    publisher    = {Zenodo},
    doi          = {10.5281/zenodo.21864536},
    url          = {https://zenodo.org/records/21864536},
}

% Appendix sections are numbered differently than article ones.
%\appendix  % Put this first.

% Can repeat for each appendix.
%\section{Appendix 1}
%Appendix text.

% You can use subsection, etc.
%\subsection{Appendix 1 Subsection}
%Appendix text.

\end{document}